\documentclass[11pt]{article}

\usepackage{amsmath,amssymb}
\usepackage[affil-it]{authblk}

\usepackage{units}
\usepackage{subfigure}
\usepackage{graphicx}

\usepackage{eurosym}

\usepackage{float}

\usepackage[hmargin=2cm,vmargin=2.5cm]{geometry}

\graphicspath{{Kuvat/}}

\newcommand{\be}{\begin{equation}}
\newcommand{\ee}{\end{equation}}

\newcommand{\bea}{\begin{eqnarray}}
\newcommand{\eea}{\end{eqnarray}}

\begin{document}

\title{
Energy optimization in ice hockey halls I. The system COP as a multivariable function, brine and design choices.
}

%
%
%

\author[1]{A. Ferrantelli
 \thanks{Electronic address: \texttt{andrea.ferrantelli@aalto.fi}; Corresponding author}}
 
 \author[2]{P. Melois}
 
 \author[1]{M. R\"aikk\"onen}
 
  \author[1]{M. Viljanen}

\affil[1]{Aalto University, Department of Civil and Structural Engineering,
FI-00076 Aalto, Finland}
\affil[2]{
Institut National des Sciences Appliqu\'ees de Toulouse\\
 31077 Toulouse Cedex 4, France
}

\date{\today}


\maketitle

\setcounter{table}{0}

\begin{abstract}
	
	This work is the first in a series of articles addressing the energy optimization in ice hockey halls.
	Here we adopt an analytical method, called functional optimization, to find which design and operating conditions maximize the Coefficient Of Performance of the entire cooling system (brine pumps and cooling tower), which we call ${\rm COP}_{sys}$.
	
	${\rm COP}_{sys}$ is addressed as a function of several variables, like electric consumption and brine physical properties. By maximizing such function, the best configuration and brine choices for the system can thus be determined accurately and rigorously.
	
	 We investigate the importance of pipe diameter, depth and brine type (ethylene glycol and ammonia) for average-sized ice rinks. An optimal brine density is found, and we compute the weight of the electric consumption of the brine pumps on ${\rm COP}_{sys}$.
	 
	Our formulas are validated with heat flow measurement data obtained at an ice hockey hall in Finland. They are also confronted with technical and cost-related constraints, and implemented by simulations with the program COMSOL Multiphysics.
	
	The multivariable approach here discussed is general, and can be applied to the rigorous preliminary study of diverse situations in building physics and in many other areas of interest.

\end{abstract}

\maketitle

\newpage



\section{Introduction}\label{sec:intro}

	Indoor ice hockey halls are large buildings which tend to be affected by very high energy consumption, due to refrigeration, heating and air conditioning. The yearly demand of an average ice hall is of the order $\sim 1800\,MWh$. Clearly, reducing this important amount of energy as much as possible is a key issue for sustainability. 
	%
	%
	%
	%
	Such a task can be accomplished in several ways, for instance via mathematical studies of the freezing process \cite{conform,conform2,heattransfer,unk141,ventilation,heattransfer2,heattransfer3}.
	%
	
	 Here we adopt a complementary approach toward energy saving, that is already employed in several fields\footnote{For example, in the theory of control and neural networks \cite{functional1,functional2,functional4,functional6,functional5,functional7,functional3,functional8}.} under the name Functional Optimization (FO). The basic idea is the following: focus on the quantities of interest and investigate their analytical behavior in function of design parameters. This gives an optimal theoretical configuration (see Sections \ref{sect:cop} and \ref{sect:brine}), which is then constrained by design and economical limits (Sections \ref{sect:tech} and \ref{sect:costs}). Eventually, one obtains practical indications on how the system could be improved (e.g. best brine, pipe size etc.),
	 %
	by means of a simple analytical calculation, that is a maxima/minima study of a multivariable function
.

This is useful for assessing energy saving already in existing ice rink arenas, e.g. in determining the more performing secondary coolant. Furthermore, the function which is studied could be \emph{any} quantity of interest to a specific assessment. Thus the method here addressed is so general that it can be systematically used in studies of energy-efficiency as a whole, not only in engineering but even in architecture.
	Just single out the quantities which are most important for energy saving, case by case, then write down their expressions in function of the system parameters, and finally study these functions to obtain the more performing configuration.
	 %
	%
	
	As a specific example, in the present article we apply FO to the Coefficient Of Performance (COP) of the \emph{entire} cooling system (refrigerator and cooling tower)\footnote{Differently from usual methods, we do not consider the COP of the compressor alone, because the bound on the evaporation temperature in ice hockey halls ($\sim -15\,^oC$) would be too restrictive.} of an ice hall located in Lepp\"avaara (Espoo, Finland).
	The COP is written in function of the system variables as $\rm{COP}_{sys}$, by using both theoretical formulas and experimental data of the refrigeration system at the ice hall in standard operating conditions. The according expression is thus automatically validated.

	We find first of all that technically acceptable values of the volumetric flow do not affect appreciably the system COP. We show indeed that the pumping power required to overcome the pressure drop in the two ice tracks, accounts for only $\sim 7\%$ of the total electric consumption. The brine pumps, the chiller, compressor and condensers are instead much more relevant, since they correspond to the remaining $93\%$. Thus optimization of the COP cannot be achieved by acting only on the pumping power (that is, by changing \emph{only} the volumetric flow).

	%
	%
%
	%
	%
	%
%
	 %
%
%

	Choosing an optimal brine fluid (the secondary refrigerant) is instead crucial to achieve a higher COP. In Section \ref{sect:brine} we show indeed that $\rm{COP}_{sys}$, through the cooling power, is particularly sensitive to the \emph{specific heat} of the fluid. Moreover, we find that it is maximal for a critical value of the density, independently of the specific brine chosen.
By means of these results, we compare ammonia and ethylene glycol, two of the most common secondary coolants. We find that ammonia gives a higher COP than ethylene glycol, the latter to be preferred in any case at concentration between 20\% and 34\% (Section \ref{sect:costs}).
	
	We also find that the pipe size and depth inside the concrete slab are instead irrelevant to the COP. However, by performing simulations with COMSOL Multiphysics, we show that they certainly cover a role in the general cooling process.
	
	First of all, we confirm rigorously the well-known practical result that the pipe size and depth do not enhance the process appreciably.
Nevertheless, increasing the pipe number by 1/3 (from 150 to 200) provides with a more uniform temperature profile at the ice surface (Sect.\ref{sect:tech}). Even taking into account the according price increase, this might be a viable option in the context of control study of the cooling process (Section \ref{sect:costs}).

	%
	%
	%
	%

\section{Ice rink, refrigeration system and measurements}\label{sect:system}

In the following, we develop a theoretical model based on measurements taken at an ice hockey hall built in 2009 in Lepp\"{a}vaara (Espoo, Finland), that contains two identical ice rinks with dimensions 28mx58m.
Heat flux and temperature are measured with a plate installed at the ice-concrete slab interface.
%
%
%
%
%
	
\begin{figure}[t]
	\centering
\begin{minipage}[b]{0.5\linewidth}
\centering
\includegraphics[width=0.7\textwidth]{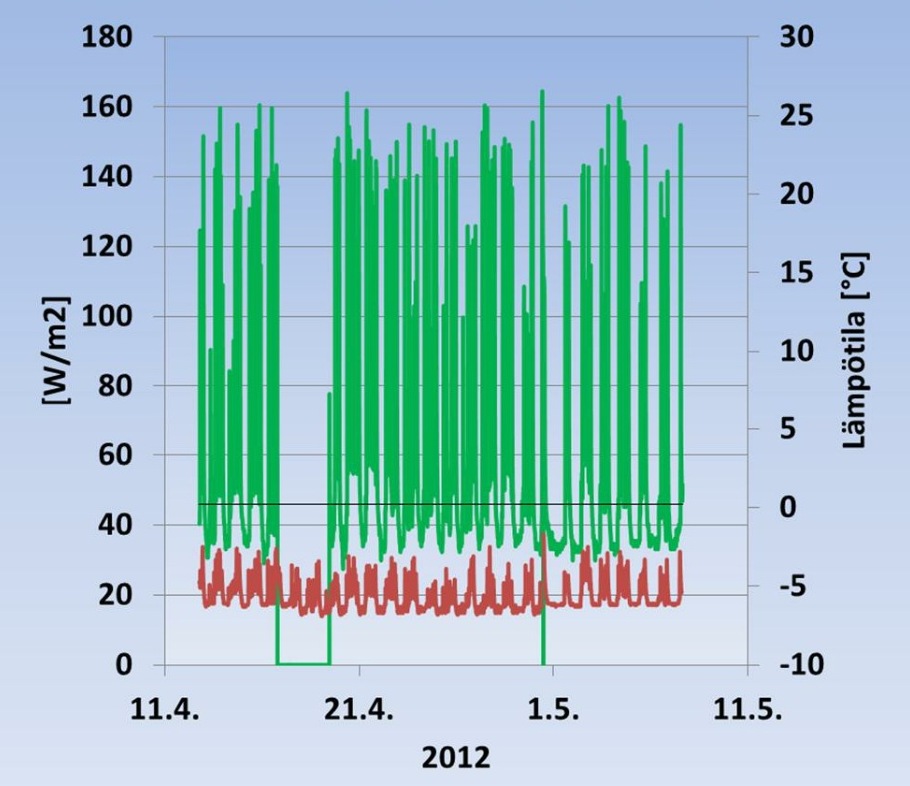}
\caption{Heat flux and interface temperature in the ice track n.1 in the Lepp\"{a}vaara ice hockey hall \cite{miska}.}
\label{fig:qTlepp}
\end{minipage}
\hspace{0.5cm}
\begin{minipage}[b]{0.45\linewidth}
\centering
\includegraphics[width=\textwidth]{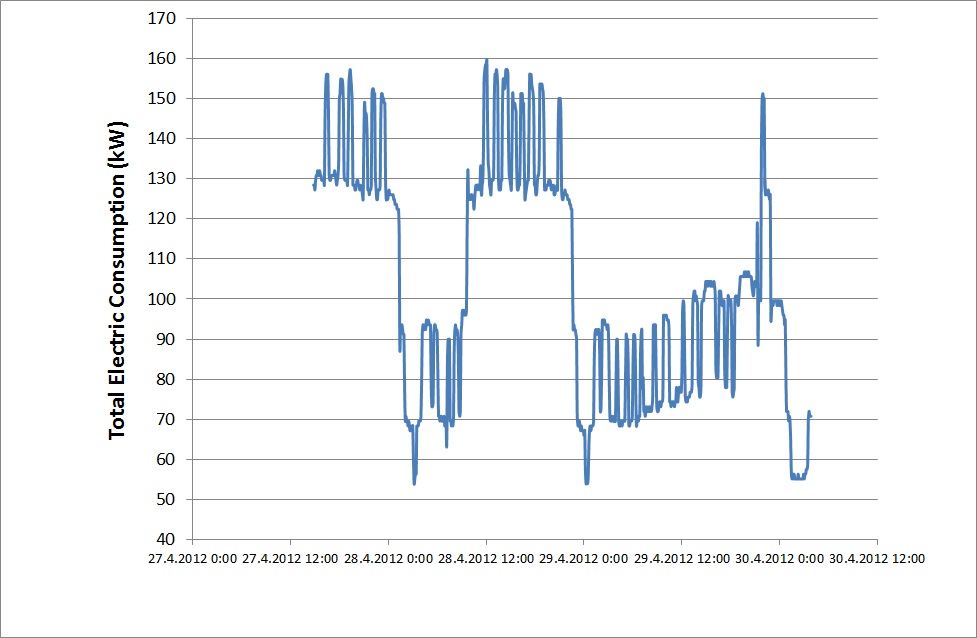}
\caption{Total electric consumption $E_{tot}=W_p+E$, as measured in four days.}
\label{fig:totalcons}
\end{minipage}
\end{figure}

	The heat flux plate and thermal sensors are connected to a logger, which records an input sent by the sensors every five minutes. Thus we get instantaneous values of the interface temperature in [C], and of the heat flux on the ice pad in $[W/m^2]$, as in Fig.\ref{fig:qTlepp}.

	%
	%
	The refrigerating machine is located in a separate container, in a backyard of the main building. The electric consumption of the whole refrigerating machine including compressors, brine pumps, evaporators and condensers is measured by a pulse meter, which is connected to a logger set to save the cumulative value of the pulses received every five minutes (Fig.\ref{fig:totalcons}).

	%
	%
%
	%
	%
	The design power of the refrigerating machine is 610 kW. Refrigeration is guided by the thermal sensors of the ice track. The evaporation temperature $T_e$ is $-15 ^oC$ and the condensation temperature $T_c$ is between $24 ^oC$ and $26 ^oC$ in springtime, $\sim 34 ^oC$ during the summer. $T_c$ is indeed controlled by the outdoor temperature. There are three compressors, and the circulating refrigerant is ammonia ($NH_3$). 
	
	The brine used is an ethylene glycol solution at 32\%, and the power of the brine pump for the condensate is 7.5 kW. There are 10 condensers \cite{Melinder}, and the power of the condenser fan is 3.2 kW. Each pipe has diameter d=22mm and runs along the 60 meters-long side of the ice track. It is connected to both the inlet and the outlet collector pipes, with D=10cm, for a total length of 120m.

\newpage


\section{The COP as a multi-variable function}\label{sect:cop}

\begin{figure}[t]
	\centering
\begin{minipage}[b]{0.5\linewidth}
\centering
\includegraphics[width=0.7\textwidth]{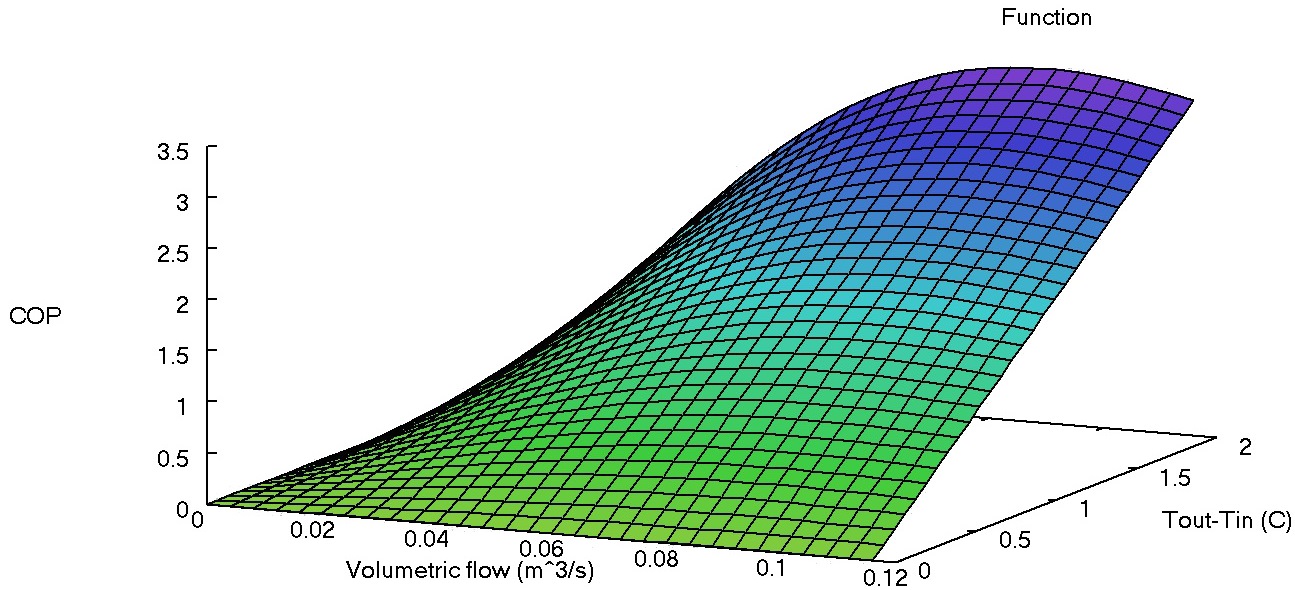}
\caption{$\rm{COP}_{sys}$ as a function of $\Delta T$ and $\dot{V}$.}
\label{fig:cop2D}
\end{minipage}
\hspace{0.5cm}
\begin{minipage}[b]{0.45\linewidth}
\centering
\includegraphics[width=\textwidth]{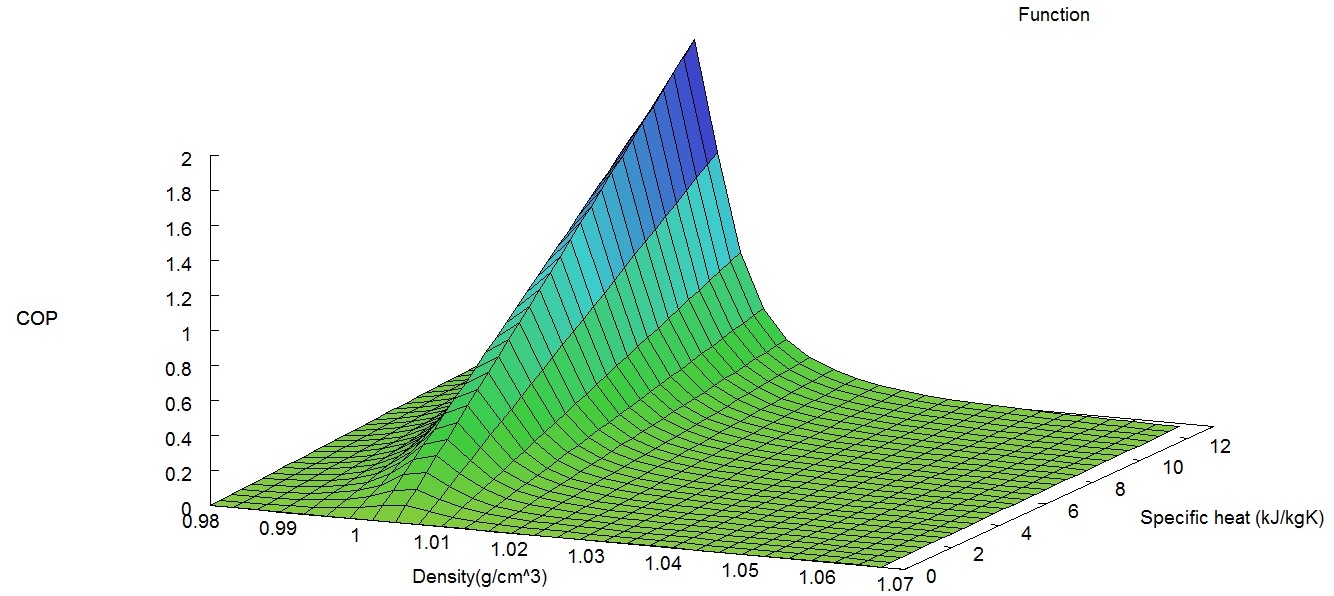}
\caption{$\rm{COP}_{sys}$ as a function of $\rho$ and $c_p$.}
\label{fig:coprhocp}
\end{minipage}
\end{figure}

Optimization of a cooling system can be achieved in several ways. Here we concentrate on maximizing the Coefficient Of Performance (COP) of the \emph{entire} cooling system. It is defined as
\be
{\rm COP}_{sys} = \frac{{\rm cooling\, power}}{{\rm electric\,consumption}}=
\frac{Q}{W_p+E}\,,
\ee
where we have split the energy consumption in two contributions: the pumping power $W_p$ required by the brine pumps for the two ice tracks, and the electric consumption $E$. The latter comprises the power consumed by the refrigeration system described in Section \ref{sect:system}, namely brine pumps, compressors, condensers and evaporators.
$W_p$ and the cooling power $Q$ are functions of the volumetric flow $\dot{V}$, brine temperature difference, geometry of the cooling system (length and pipe diameters) and brine physical properties.
%
%
%
%

	Let us now expand the expressions for the cooling power $Q$ and for the pumping power $W_p$, to rewrite the COP in function of the variables of interest. First, expand the cooling power:
\be\label{COP}
{\rm COP}_{sys}=\frac{Q}{W_p+E}=\frac{\dot{m}c_p\Delta T}{W_p+E}=\frac{c_p\rho\dot{V}\Delta T}{W_p+E}\,.
\ee
Then, the total pressure drop is the sum of contributions from the collector pipes and from the pipes under the ice rink (subscript $p$). The pumping power can thus be written in the following form,
\be\label{Wp}
W_{p_i}=\frac{\dot{V}_i}{\eta}\Delta p_i = \frac{\dot{V}_i}{\eta}\left( f\frac{L}{D} \frac{\rho}{2} v_{m_i}^2 + f_p\frac{l}{d} \frac{\rho}{2} v_{m_i,p}^2 \right)=\frac{0.14}{\eta}\rho\dot{V}_i^2\left( \frac{L\nu^{1/5}}{D^{24/5}\dot{V}_i^{4/5}}+1.91\frac{l}{d^4}\nu \right)\,,
\ee
where the index $i=1,2$ labels the two ice tracks and $\eta=0.75$ is the efficiency of each brine pump. $L$ and $D$ are the length and diameter of the main collector pipes with cross-sectional area $A$, $l$ and $d$ are those of the 150 pipes under each ice track, with cross-sectional area $A_p$. The mean bulk velocities of the brine in the collector and ice rink pipes are respectively $v_{m_i}=\dot{V}_i/A$ and $v_{m_i,p}$. The pipes in the concrete slab are connected in parallel, so $\dot{m}_i=150\,\dot{m}_{i,p}$ implies $v_{m_i,p}=\dot{V}_{i_p}/A_p=\dot{V}_i/(150 A_p)$.

	$f$ and $f_p$ are the corresponding friction coefficients. It can be shown that, for our average volumetric flow $\dot{V}= 28\,l/s$, the fluid motion in the collector pipes is turbulent (${\rm Re}\sim 32400$), while in the smaller pipes it is laminar (${\rm Re} = 982$). This justifies the exponentials in Eq.(\ref{Wp}). The minor losses from the pipe elbows are negligible.

	By substituting the above expression for the pumping power in (\ref{COP}) and simplifying, we get our formula:
	\be\label{copmaster}
	{\rm {\rm COP}_{sys}} = 
	\frac{\rho c_p(\dot{V}_1\Delta T_1+\dot{V}_2\Delta T_2)}
	{W_{p_1}+W_{p_2}+E}=
	\frac{c_p(\dot{V}_1\Delta T_1+\dot{V}_2\Delta T_2)}
	{\dfrac{1}{\eta}\left[ 0.14\dfrac{\nu^{0.2}}{D^{4.8}}L(\dot{V}_1^{2.8}+\dot{V}_2^{2.8})+0.27\dfrac{\nu}{d^4}l(\dot{V}_1^2+\dot{V}_2^2)\right]+\dfrac{E}{\rho}}\,.
	\ee
	Therefore one can express the COP implicitly as follows,
\be\label{fcop}
COP = f(\dot{V},\rho,\nu,c_p,L,d,E)\,,
\ee
namely by considering the COP as a function of the above variables, one could investigate the impact of each of them on the energy optimization, as in Figs.\ref{fig:cop2D} and \ref{fig:coprhocp}.
	Finding for which values of these parameters the function COP is maximal corresponds to choosing the best possible configuration of brine fluid, volumetric flow etc. We are therefore applying to our problem the method of \textsl{Functional Optimization} (FO), which is used in a number of diverse research topics (see e.g. \cite{functional1,functional2,functional4,functional6,functional5,functional7,functional3,functional8}).

	In the next sections, we will take into consideration each and all of the variables entering Eq.(\ref{copmaster}), and calculate for which of their values the function in Eq.(\ref{fcop}) is maximal.


	\subsection{\normalsize Relative weight of pumping power and electric consumption on ${\rm COP}_{sys}$}\label{sect:pump}

\begin{figure}[t]
\centering
\begin{minipage}[b]{0.50\linewidth}
\centering
\includegraphics[width=\textwidth]{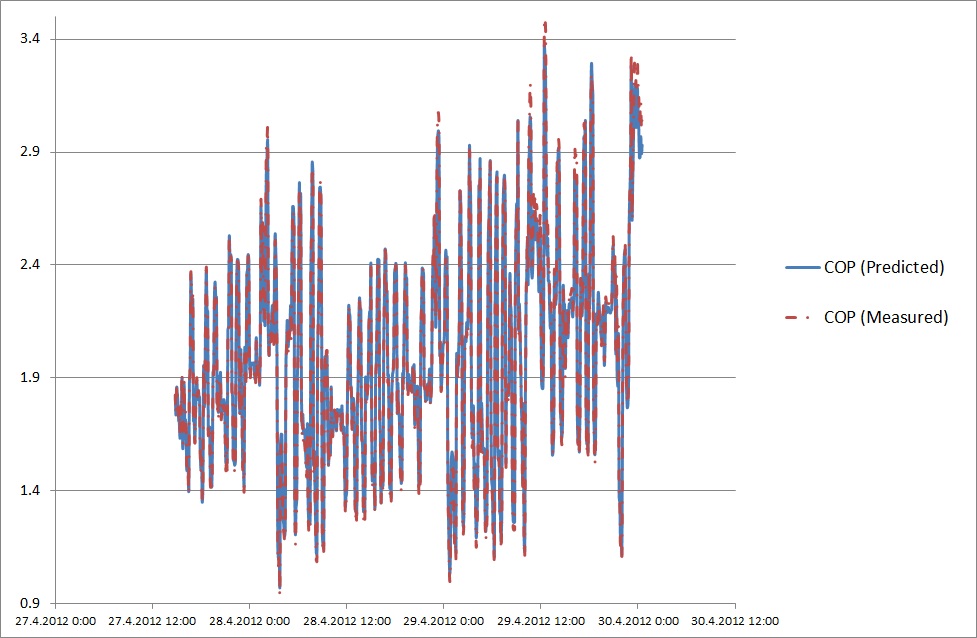}
\caption{COP computed with Eq.(\ref{copmaster}), vs the COP as measured at the ice rink.}
\label{fig:validation}
\end{minipage}
\hspace{0.5cm}
\begin{minipage}[b]{0.45\linewidth}
\centering
\includegraphics[width=\textwidth]{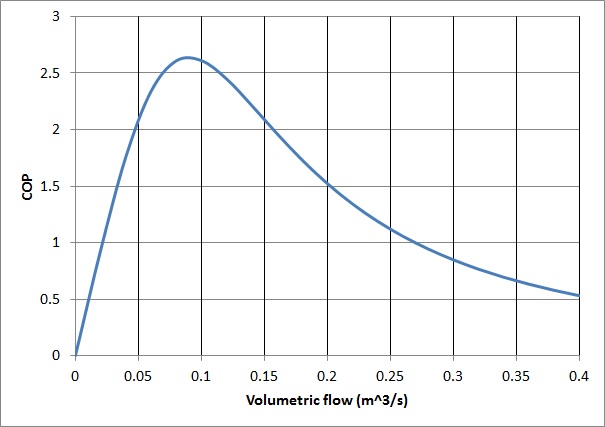}
\caption{${\rm COP}_{sys}$ as a function of only the volumetric flow.}
\label{fig:COPV}
\end{minipage}
\end{figure}

	We first focus on the relative impact of $W_p$ and $E$ on the COP. Eq.(\ref{copmaster}) follows from the theory, but it relies on the experimental data as well. First of all, the volumetric flow for the ice track $i$ is $\dot{V}_i=\dot{V}\eta_v$, where $\eta_v$ is the speed control of the according brine pump, $0<\eta_v<1$, and $\dot{V}=28\,l/s$ is the maximum volumetric flow. $\eta_v$ and $E_{tot}=W_p+E$ are measured every 5 minutes, giving directly $\rm{COP}_{sys}$ as in Fig.\ref{fig:validation}.
	In Eq.(\ref{COP}), the denominator $W_p+E$ corresponds to the \textit{total} electric consumption of the compressor,
	\be
	E_{tot}=W_{p_1}+W_{p_2}+E=W_p+E\,,
	\ee
	that is measured at the ice rink in Lepp\"avaara. This is, on the average, of order $E_{tot}\sim 110 kW$. Since our data do not contain values of the pumping power $W_p$ and $E$ separately, we compute explicitly $W_p$ and quantify its impact on the energetic consumption of the system. We find an average $W_p=8.61\,kW$, corresponding to a weight of $6.6\%$.
	
	%
	This confirms a well-known result, and implies that $E=P_1+ P2+E_{cooling}\sim 93.4\%$ (see for instance \cite{encons}). $P_i$ is the power of the brine pump $i$, valued between 10kW and 15kW depending on the volumetric flow. $E_{cooling}\sim 76.4\,kW$ is instead the consumption of the condensers, chiller and compressor, and it is therefore the largest contribution to $E_{tot}$.
	%
	%
	%
%
%

	Consider now Eq.(\ref{copmaster}) as a function of only the volumetric flow $\dot{V}$ and of $\Delta T$, the temperature difference of the brine at inlet and outlet.
	%
%
				%
	%
	%
	%
	${\rm COP}_{sys}$ is clearly proportional to $\Delta T$, whereas the dependence on the volumetric flow is nontrivial ($E$ is independent of $\dot{V}$).

		The COP can be plotted in a 2-dimensional graph	as in Figures \ref{fig:cop2D} and \ref{fig:coprhocp}, or as a function of only $\dot{V}$ if $\Delta T$ is fixed. We show this for $\Delta T=1.5^oC$ in Figure \ref{fig:COPV}, where one can see that the COP increases until it reaches a maximum for a specific value of the volumetric flow, before decreasing again.
		The maximum corresponds to $\dot{V}\sim 90\,l/s$, that is a very large value. In other words, at least in theory it is possible to reach the maximal ${\rm COP}_{sys}$ by increasing the pumping power. This is expected, the cooling power is $\propto \dot{V}$ and the pumping power $W_p$ weights for only $\sim 7\%$ of the total electric consumption.

		Taking into account all the results of this section, we conclude that the brine pumps are not determinant for energy optimization.



		\section{Brine choices}\label{sect:brine}

		First of all, recall Eq.(\ref{copmaster}), namely
			\be
	{\rm COP}_{sys} = \frac{c_p\dot{V}\Delta T}
	{\dfrac{1}{\eta_v}\left( 0.14\dfrac{L}{D^{4.8}}\nu^{0.2}\dot{V}^{2.8}+0.27\dfrac{l}{d^4}\nu\dot{V}^2 \right)+\dfrac{E}{\rho}}\,.\nonumber
	\ee
		 where $\dot{V}\Delta T\equiv(\dot{V}_1\Delta T_1+\dot{V}_2\Delta T_2)$ and $\dot{V}^a\equiv\dot{V}_1^a+\dot{V}_2^a$. By using the above, let us study how the brine physical characteristics are relevant for ${\rm COP}_{sys}$.\\
		We have shown that the pumping power counts for only $\sim 7\%$ of the total electric consumption, thus the kinematic viscosity $\nu$ is irrelevant. One should then focus on the density $\rho$ and the specific heat $c_p$. Since 
		\be
		{\rm COP}_{sys}\propto c_p\,,\, {\rm COP}_{sys}\propto\frac{\rho}{E}\,,
		\ee
		and in general $c_p\propto 1/\rho$, it is not trivial to decide which parameter should be increased to achieve higher COP. One then needs to write $\nu$ and $c_p$ in function of $\rho$, to obtain a plot ${\rm COP}_{sys}$ vs $\rho$ which singles out the point of maximum.

\begin{figure}[t]
\centering
\begin{minipage}[b]{0.45\linewidth}
\centering
\includegraphics[width=\textwidth]{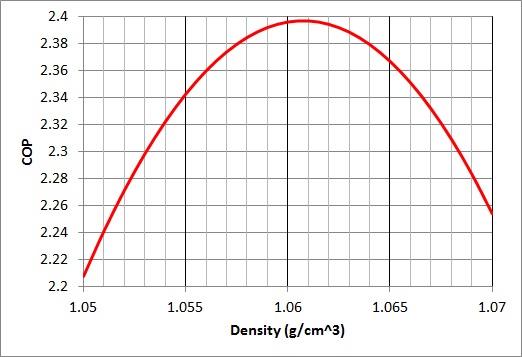}
\caption{$\rm{COP}_{sys}$ in function of the brine density $\rho$.}
\label{fig:copvsrho}
\end{minipage}
\hspace{0.5cm}
\begin{minipage}[b]{0.45\linewidth}
\centering
\includegraphics[width=\textwidth]{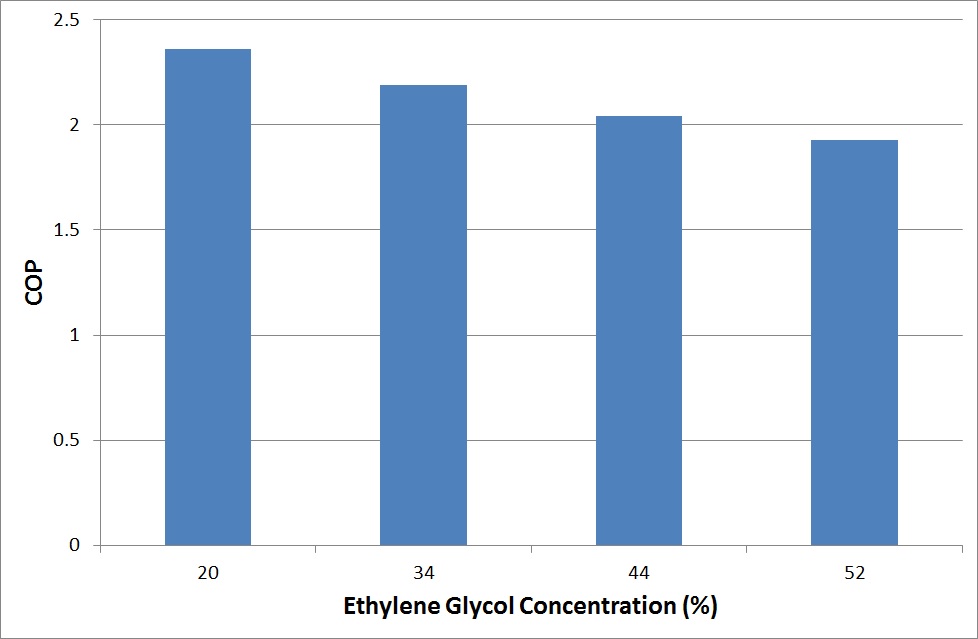}
\caption{${\rm COP}_{sys}$ as a function of the ethylene glycol concentration.}
\label{fig:copvsc}
\end{minipage}
\end{figure}

		However, there are no theoretical formulas $\nu=f(\rho)\,,c_p=f(\rho)$, therefore we interpolate the experimental data taken at our laboratory, for an aqueous solution of ethylene glycol at 34\%.
We obtain the following relations,
		\be
		\nu(\rho) =24725\rho^2-524575\rho+278244\,,\qquad
		c_p(\rho) = -2517.5\rho^2+5337.6\rho-2825.54\,.
		\ee
	These provide with estimates which are sufficiently accurate for our purpose.
		 For typical values ($\dot{V}_1\sim 20\, l/s$, $\dot{V}_2\sim 28\, l/s$, $\Delta T_i\sim 1.5^oC$, $E\sim 110\,kW$), the COP is then a function of the brine density only, and the optimal value is found to be $\rho=1.061\,g/cm^3$, see Fig.\ref{fig:copvsrho}.

		 Last, we study how ${\rm COP}_{sys}$ changes according to the average brine temperature $(T_{in}+T_{out})/2$. We use again our experimental data to derive the following interpolation formulas for $\rho(T)$, $\nu(T)$, $c_p(T)$,
		\be
		\rho= 1.1176-0.0002T\,,\;
		\nu= 0.01T^2-5.678T+810.08\,,\;
		c_p= -0.0001T^2+0.0575T-4.6061\,.
		\ee
	By substituting these expressions into Eq.(\ref{copmaster}), we obtain the ${\rm COP}_{sys}$ values in function of the average brine temperature and find agreement with the measurements.
		%
		%
		%
		%
%
		%
		%
		%
		%
		%
	To generalize these results, we compare the COP obtained for four different ethylene glycol concentrations, with the same system configuration (volumetric flow, electric consumption etc.).

		%
		%
			%
	%
	%
		%
		%
		It is shown in Figure \ref{fig:copvsc} that the lower the concentration, the higher ${\rm COP}_{sys}$.
		%
		%
		Moreover, at fixed $T=-10^oC$ both $\nu$ and $\rho$ increase, while $c_p$ decreases. Since ${\rm COP}_{sys}$ eventually behaves like $c_p$, we can say that for a common (ethylene glycol) brine, \emph{the specific heat is the most crucial physical parameter for optimizing the system COP}.
		%
		%
		%
		%
		%
		%
		%
		%
		%
		%
		%
		%
		
		One could also estimate which brine would be better for a higher ${\rm COP}_{sys}$, i.e. for a more efficient cooling system. We consider an aqueous solution of 14\% ammonia ($NH_3$) and calculate the according COP with (\ref{copmaster}); the result is then compared to the COP with ethylene glycol ($C_2H_6O_2$) at $20\%$ and $34\%$. We find the following,
		\bea
		&& {\rm COP}_{sys}(NH_3)>{\rm COP}_{sys}(C_2H_6O_2)\,,
		\eea
		for any ethylene glycol concentration. In particular, for a solution at 20\% the gain is 9\%, whilst for a solution at 34\% the COP obtained with ammonia is larger by $\sim$18\%, that is quite remarkable. We conclude that according to our analysis, \emph{ammonia should be preferred to ethylene glycol}.



	\section{Design constraints and cost estimate}
	
\begin{figure}[t]
		\centering
\begin{minipage}[b]{0.4\linewidth}
\centering
\includegraphics[width=\textwidth]{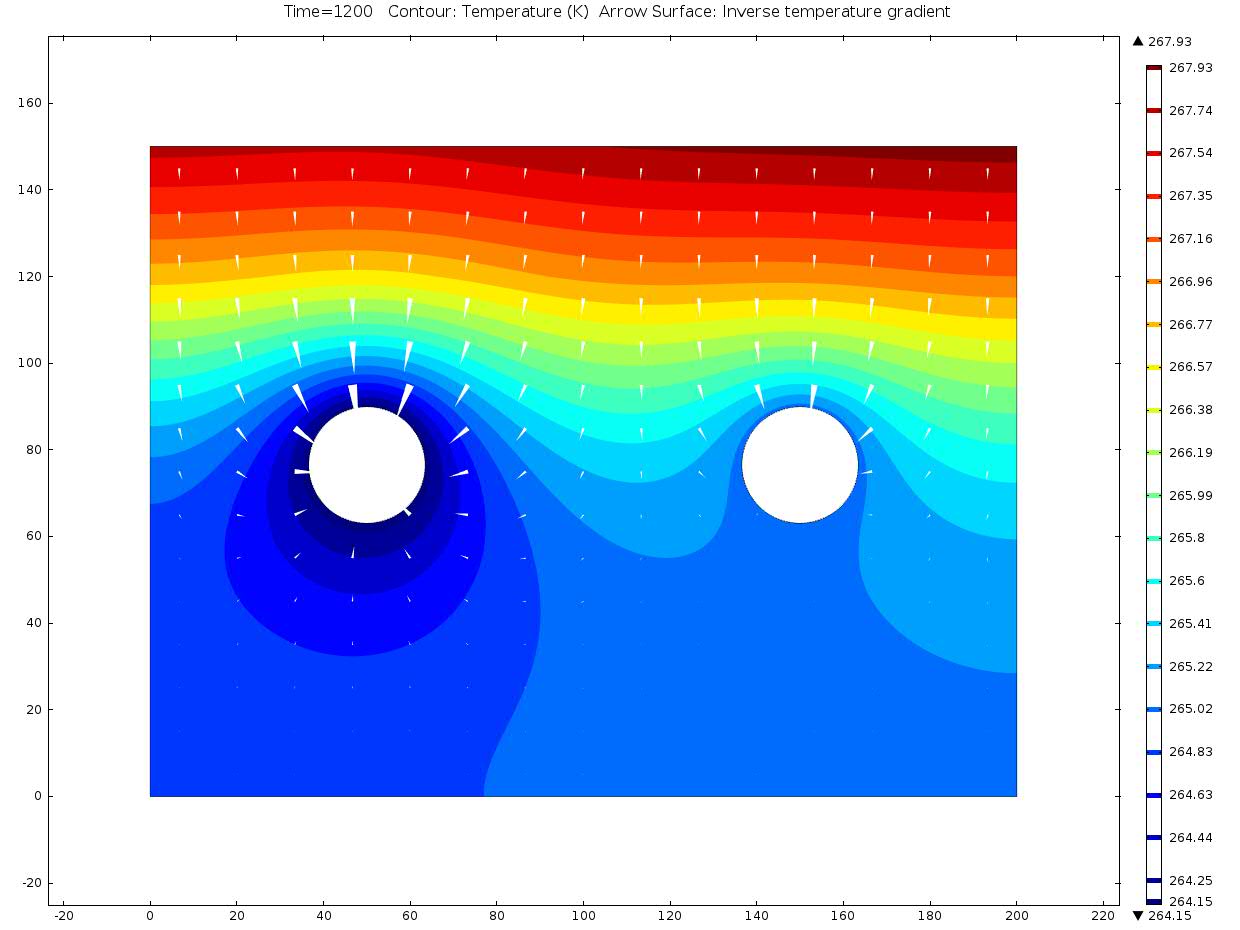}
\caption{Simulation of heat transfer in the ice/concrete slab, in the case of two pipes.}
\label{fig:comsol_2pipes}
\end{minipage}
\hspace{1.8cm}
\begin{minipage}[b]{0.4\linewidth}
\centering
\includegraphics[width=\textwidth]{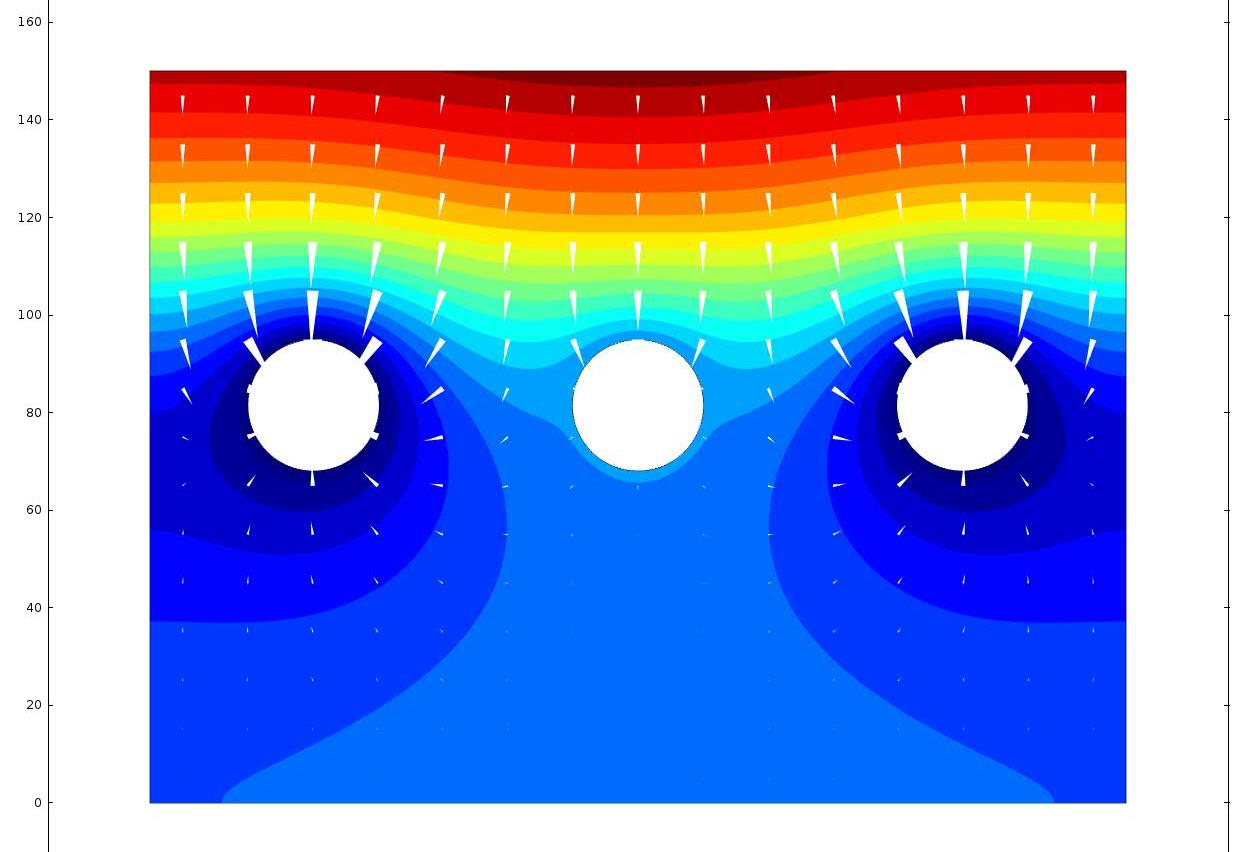}
\caption{Simulation of heat transfer in the ice/concrete slab, in the case of three pipes.}
\label{fig:comsol_3pipes}
\end{minipage}
\end{figure}

	We now address the question of how are the pipe number, pipe diameter and pipe depth in the slab relevant to the heat transfer.
		We consider a two-dimensional slab. Following its symmetry, we focus on a 200 mm wide section, containing two pipes with 100 mm separation, and perform simulations of conductive transient heat transfer with the program COMSOL Multiphysics (Figures \ref{fig:comsol_2pipes} and \ref{fig:comsol_3pipes}).

		The concrete slab has dimensions 200x120 mm, the pipe's external diameter is 27 mm, the reference depth (at the pipe's edge) is 30 mm and the ice thickness is 30 mm as well. We use densities $\rho=2300\,kg/m^3$ for concrete, $\rho=918.9\,kg/m^3$ for ice and $\rho=1055\,kg/m^3$ for a 32\% ethylene glycol solution.

		\subsection{Design constraints}\label{sect:tech}
		A uniform heat flow $q=60\,W/m^2$, consistent with our measurements at Lepp\"avaara, is applied to the ice surface, at initial temperature $T_i=-4^oC$. The inlet and outlet pipes are at temperatures $T_{in}=-9^oC$ and $T_{out}=-8^oC$.
	%
	%
	%
	%
	%

	Considering now the pipe sizes $d=27\,mm$,  $d=35\,mm$, $d=40\,mm$,
	%
we find that, for an increase in the diameter from 27 to (oversized) 40 mm, namely 148\%, the temperature changes by only $\sim 1\%$. Moreover, the temperature profiles do not change qualitatively
	. 
	We conclude that the \emph{pipe size has a very small role} in the optimization of the heat transfer. This is clearly due to the high thermal mass of ice and concrete.
	
	%
%
	%
	%
	Changing the pipe depth in the range $25\,mm<h<30\,mm$
	shows that neither in this case the system is very responsive to a large increase of the parameter (here, the depth).
	It can be shown indeed that pipes closer to ice by 20\% lower the surface temperature by only $0.1\,K$.
	
\begin{figure}[t]
		\centering
\begin{minipage}[b]{0.4\linewidth}
\centering
\includegraphics[width=\textwidth]{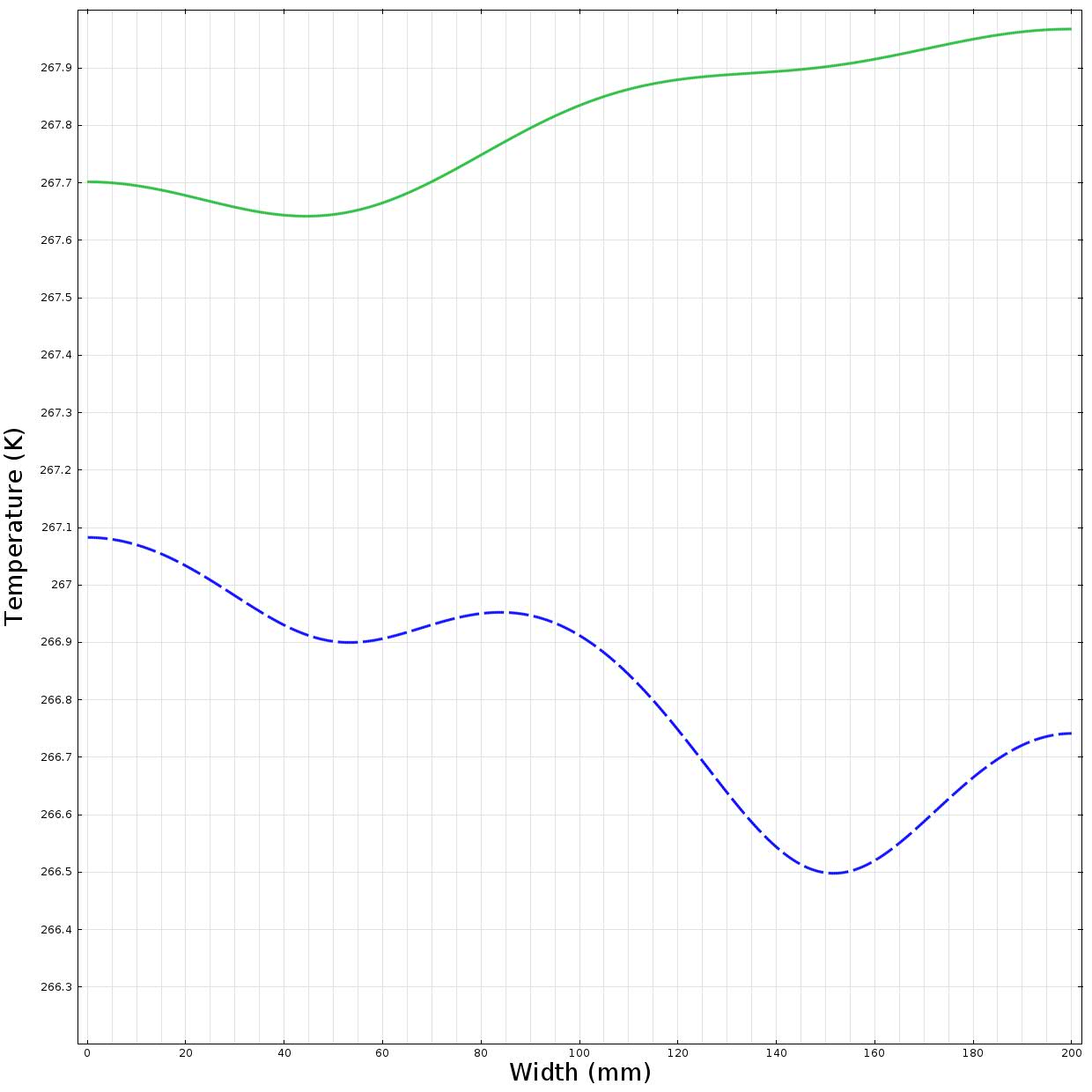}
\caption{Temperature profile at the ice surface (solid) and ice-concrete interface (dashed), in the case of two pipes.}
\label{fig:T2pipes}
\end{minipage}
\hspace{0.2cm}
\begin{minipage}[b]{0.4\linewidth}
\centering
\includegraphics[width=\textwidth]{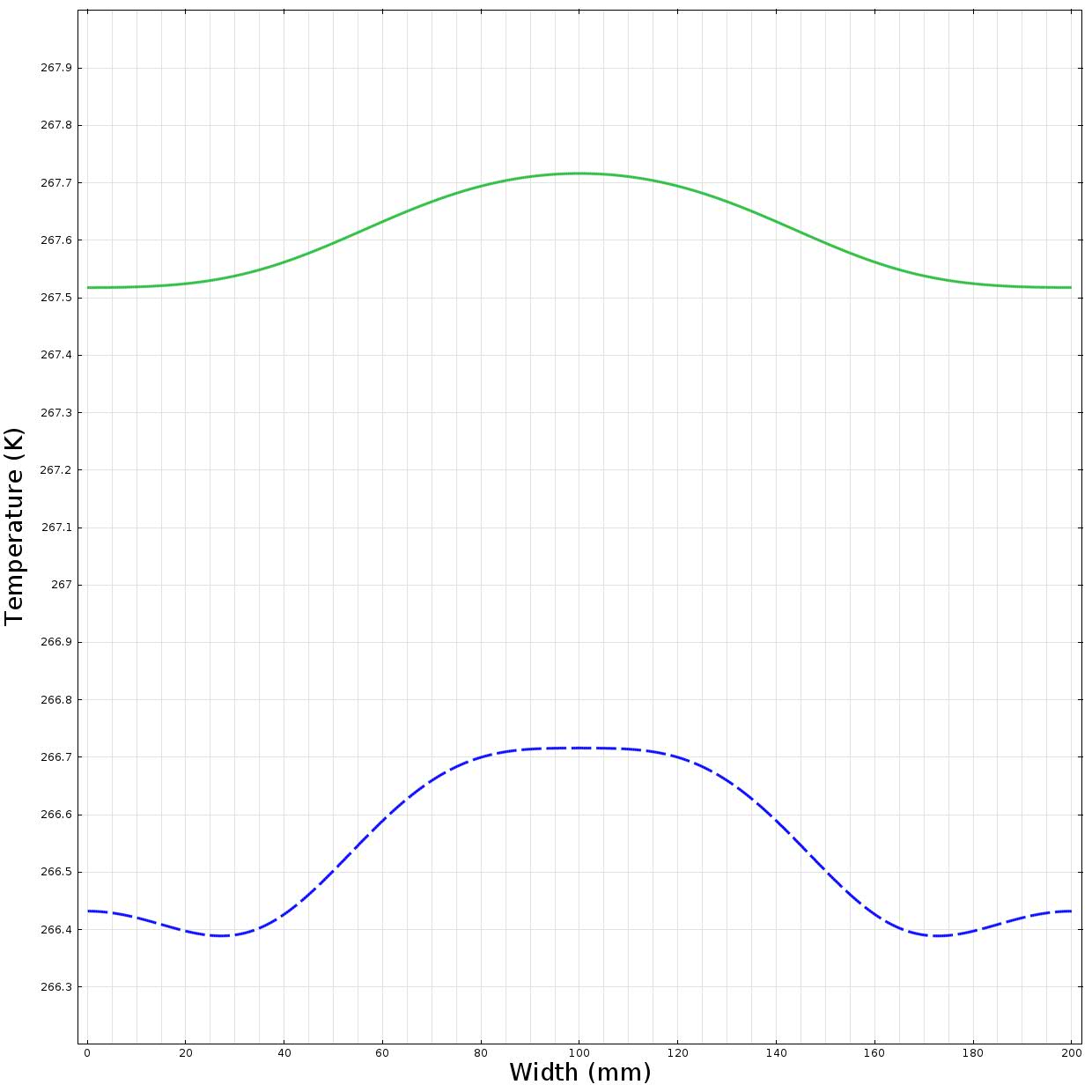}
\caption{Temperature profile at the ice surface (solid) and ice-concrete interface (dashed), in the case of three pipes.}
\label{fig:T3pipes}
\end{minipage}
\end{figure}

	%
	%
	%
	%
	%
	%
	Finally, by adding one more inlet pipe in the same width, i.e. from 2 to 3 pipes every 200 mm,
	%
	the temperature profiles at the ice surface and ice/concrete interface are radically different in the two cases, see Figs.\ref{fig:T2pipes} and \ref{fig:T3pipes}.	
The temperature at $x=200\,mm$ is almost 0.5$^oC$ lower for three pipes, and its profile is much more uniform.
	If the smaller distance of 66mm does not constitute a problem for the structure, using 200 pipes instead of 150 for the whole ice rink might then be an interesting option.




	\subsection{\normalsize A preliminary cost estimate}\label{sect:costs}

The top-down approach described in the previous section has been validated with measurements performed at an ice hockey arena.
	This section aims to complement our results with a bottom-up approach, discussing their practical limits \cite{unk61}
	.
%
%
%
%

	The refrigerating slabs contain pipes buried in concrete, and there are several limitations on the thickness of the concrete layer. 
%
In our case, the maximum load is inferior to $10\,kN/m^{2}$ (distributed or concentrated).
	For an 80mm to 100mm polystyrene insulation layer (that can resist to maximum $850\,kN/m^{2}$), the minimum thickness of the concrete slab should be 65mm to resist to the maximum vertical load. Usually, this size goes up to 120mm in order to accommodate the pipes.
	Overall, these limitations lead to design the concrete layer with a minimum concrete thickness of 25mm above the edges of the pipes.
	%
		%
	%
	%
	%
	
	The study performed in this paper is based on a 200mm wide slab containing 2 or 3 pipes spaced out 100mm or 66mm each. Let us consider an Olympic ice rink of 30m width, for a total of 300 or 450 pipes. Then the total length of the PVC pipes would be 17\,000m or 27\,000m (60m per pipe).
%
%
%
%
%
	
	We find that placing 3 pipes rather than 2 inside a 200mm slab demands an additional expense of more than $50\%$. However, we will consider this option in a future work in the context of control, to determine whether the higher installation price is balanced by higher performance.
	%
	%
	%
	%
	Last, the price of liquid ammonia seems to be lower than the price of ethylene glycol.
%
%
%
%
%
%
%
%
	
	For a -8$^\circ C$ freezing point (average pipe temperature in our data) the minimal concentration is 18,1\%vol. This increases to 52\%vol for -39$^\circ C$~\cite{Concentrations2}. Moreover glycol concentrations less than 20\%vol becomes a nutriment source for bacteria~\cite{Yuzwa}.
	To sum up, ideal ethylene glycol concentrations should be between 20 and 52\%vol, and in any case, liquid ammonia seems to be \emph{the best solution}.
	%



\section{Conclusions and outlook}

	In this paper we have discussed a functional method to study quantities of technological interest, and applied it to the optimization of cooling in ice hockey halls. In this specific case, we have provided with considerations about energy consumption and suggestions for the optimal brine and piping design.
%
	%
	 %
	 %
	
	We find that the pumping power $W_p$ corresponds to only $\sim 7\%$ of the total electric consumption, and that the system Coefficient Of Performance ${\rm COP}_{sys}$ is not very sensitive to the volumetric flow. The brine pumps, chiller, compressor and condensers account instead for the remaining $\sim 93\%$. We also show that choosing an optimal secondary coolant is determinant for a higher ${\rm COP}_{sys}$, which is particularly sensitive to the \textsl{specific heat} of the brine. Moreover, there is a specific value of the \textsl{density} for which the COP is maximal,
 independently of the specific brine chosen.

	We conclude that ammonia is more performing than ethylene glycol, and that glycol at concentration between 20\% and 34\% is to be preferred in any case. We have compared these predictions with technical constraints, and a cost estimate is provided in Section \ref{sect:costs}.

	Design possibilities regarding pipe size and depth inside the concrete slab have been investigated as well, via simulations with COMSOL Multiphysics. We confirm that such parameters do not enhance nor hinder the process appreciably. However, increasing the number of pipes provides with a more uniform temperature profile at the ice surface. This option will be considered in our forthcoming works in this series.

	 Our results are obtained with a model that is based on measurements performed at an ice rink in Finland, and are confronted with technical and cost-related constraints.
	 Investigating the COP of the \emph{entire} cooling system as a function of several variables is an interesting option,	as it provides with precise indications on the system design that help to reduce expensive and time-consuming preliminary tests. Moreover, the multivariable approach described in this paper is general, and can be applied to the rigorous study of diverse situations in engineering and architecture.

\section{Acknowledgements}

The authors would like to thank Kirsi S\"{a}kkinen for the measurements of the brine properties, and acknowledge support by the Academy of Finland.

\end{document}